\documentclass[10pt, conference, letterpaper]{IEEEtran}

\RequirePackage{graphics}
\usepackage{epsfig}
\usepackage{amsmath}
\usepackage{amsfonts}
\usepackage{graphicx,subfigure}
\usepackage{amssymb}
\usepackage{algorithm}
\usepackage{algorithmic}
\usepackage{color}
\usepackage{framed}
\usepackage{xcolor}
\usepackage{lipsum}

\newcommand{\tabincell}[2]{\begin{tabular}{@{}#1@{}}#2\end{tabular}}
\def\overbracket#1{\mathop{\vbox{\ialign{##\crcr\noalign{\kern3\p@}
\downbracketfill\crcr\noalign{\kern3\p@\nointerlineskip}
$\hfil\displaystyle{#1}\hfil$\crcr}}}\limits}
\def\underbracket#1{\mathop{\vtop{\ialign{##\crcr
$\hfil\displaystyle{#1}\hfil$\crcr\noalign{\kern3\p@\nointerlineskip}
\upbracketfill\crcr\noalign{\kern3\p@}}}}\limits}

\usepackage{cite}

\ifCLASSINFOpdf
\else
\fi

\begin{document}
\title{The Sound and the Fury: \\ Hiding Communications in Noisy Wireless Networks with Interference Uncertainty}


\author{\IEEEauthorblockN{
Zhihong Liu\IEEEauthorrefmark{1}, Jiajia
Liu\IEEEauthorrefmark{1}\IEEEauthorrefmark{2}\IEEEauthorrefmark{5}, Yong Zeng\IEEEauthorrefmark{1},
Li Yang\IEEEauthorrefmark{3}, and Jianfeng Ma\IEEEauthorrefmark{1}}
\IEEEauthorblockA{\IEEEauthorrefmark{1}School of Cyber Engineering, Xidian University, Xi'an,
China} \IEEEauthorblockA{\IEEEauthorrefmark{2}State Key Laboratory of Integrated Service Networks,
Xidian University, Xi'an, China}

\IEEEauthorblockA{\IEEEauthorrefmark{3}School of Computer Science and Technology, Xidian
University, Xi'an, China}
\IEEEauthorblockA{\IEEEauthorrefmark{5}E-mail:
liujiajia@xidian.edu.cn} }


\maketitle

\begin{abstract}
Covert communication can prevent the adversary from knowing that a wireless transmission has
occurred. In the additive white Gaussian noise channels, a square root law is obtained and the
result shows that Alice can reliably and covertly transmit $\mathcal{O}(\sqrt{n})$ bits to Bob in
$n$ channel uses. If additional ``friendly'' node near the adversary can inject artificial noise to
aid Alice in hiding her transmission attempt, covert throughput can be improved, i.e., Alice can
covertly transmit $\mathcal{O}(\min\{n,\lambda^{\alpha/2}\sqrt{n}\})$ bits to Bob over $n$ uses of
the channel ($\lambda$ is the density of friendly nodes and $\alpha$ is the path loss exponent of
wireless channels). In this paper, we consider the covert communication in a noisy wireless
network, where Bob and the adversary Willie not only experience the background noise, but also the
aggregated interference from other transmitters. Our results show that uncertainty in interference
experienced by Willie is beneficial to Alice. When the distance between Alice and Willie
$d_{a,w}=\omega(n^{\delta/4})$ ($\delta=2/\alpha$ is stability exponent), Alice can reliably and
covertly transmit $\mathcal{O}(\log_2\sqrt{n})$ bits to Bob in $n$ channel uses. Although the
covert throughput is lower than the square root law and the friendly jamming scheme, the spatial
throughput is higher. From the network perspective, the communications are hidden in ``the sound
and the fury'' of noisy wireless networks, and what Willie sees is merely a ``shadow'' wireless
network. He knows for certain that some nodes are transmitting, but he cannot catch anyone
red-handed.
\end{abstract}

\begin{IEEEkeywords}
Physical-layer Security; Covert Communication; Stochastic Geometry; Interference.
\end{IEEEkeywords}

\IEEEpeerreviewmaketitle

\section{Introduction}
Traditional cryptography methods for network security can not solve all security problems. In
wireless networks, if a user wishes to communicate covertly without being detected by other
detectors, encryption to preventing eavesdropping is not enough \cite{Hiding_Information}. Even if
a message is encrypted, the metadata, such as network traffic pattern, can reveal some sensitive
information \cite{ICCCN}. Furthermore, if the adversary cannot detect the transmission, he has no
chance to launch the ``eavesdropping and decoding'' attack even if he has boundless computing and
storage capabilities. On other occasions, such as in a battlefield, soldiers hope to hide their
tracks and communicate covertly. Another occasion, such as defeating ``Panda-Hunter'' attack
\cite{panda_hunter}, also needs to prevent the adversary from detecting the transmission behavior
of users to protect the location privacy.

Consider the scenario where a transmitter Alice would like to communicate with a receiver Bob
covertly over a wireless channel in order to not being detected by a warden Willie. In
\cite{square_law}, Bash \emph{et al.} found a square root law in additive white Gaussian noise
(AWGN) channels, that is, Alice can transmit $\mathcal{O}(\sqrt{n})$ bits reliably and covertly to
Bob over $n$ uses of wireless channels. The square root law implies pessimistically that the
asymptotic privacy rate approaches zero. If Willie does not know the time of the transmission
attempts of Alice, Alice can reliably transmit $\mathcal{O}(\min\{(n\log T(n))^{1/2}, n\})$ bits to
Bob while keeping the Willie's detector ineffective with a slotted AWGN channel model containing
$T(n)$ slots \cite{time1}. To improve the performance of covert communication, Lee \emph{et al.}
\cite{LDP1} found that, Willie has measurement uncertainty about its noise level due to the
existence of SNR wall \cite{SNR}, then they obtained an asymptotic privacy rate which approaches a
non-zero constant. Following Lee's work, He \emph{et al.} \cite{Biao_He} defined new metrics to
gauge the covertness of communication. They took the distribution of noise measurement uncertainty
into consideration. Wang \emph{et al.} \cite{Fundamental_Limits} considered the covert
communication over the discrete memoryless channels (DMC), and found that the privacy rate scales
like the square root of the blocklength. Bloch \emph{et al.} \cite{Bloch} discussed the covert
communication problem from a resolvability perspective. He developed an alternative coding scheme
such that, if the warden's channel statistics are known, on the order of $\sqrt{n}$ reliable covert
bits may be transmitted to Bob over $n$ channel uses with only on the order of $\sqrt{n}$ bits of
secret key. Soltani \emph{et al.} \cite{renewal} studied the covert communications on renewal
packet channels. They introduced some information-theoretic limits for covert communication over
packet channels where the packet timings of legitimate users are governed by a Poisson point
process.

Although the research on covert wireless communication focuses on the transmission capability, it
is quite different from the works that measure the performance of wireless networks
\cite{Gupta}\cite{Transmission_Weber}. In general, the covertness of communication is due to the
existence of noise that the adversary cannot accurately distinguish between the signal and noise.
If we can increase the measurement uncertainty of the adversary, the performance of covert
communication can be improved. Take the following occasion as an example,

``One day morning you walked in the woods. A lark with beautiful tail feathers was singing. You
closed your eyes, listening \dots Although a little breeze was rustling and tumbling in the woods,
you could still hear the sweet lark sing in the clear air of the day. All of a sudden, a crowd of
larks flew here, you was drowned in the noisy twitters \dots You no longer knew whether the lark
with beautiful tail feathers was still singing or not \dots''

Now the lark's song is submerged in the interference and is difficult to be detected. Interference
or jamming is usually considered harmful to wireless communications, but it is also a useful
security tool. Cooperative jamming is regarded as a prevalent physical-layer security approach
\cite{Physical-Layer-Security}\cite{Challenges} in wireless communication environment. Jammers
inject additional interferences when the transmitter sends messages in order to interfere the
potential eavesdroppers
\cite{Artificial_noise}\cite{Artificial_noise_Goel}\cite{jamming_barrier1}\cite{heartbeats}. Sobers
\emph{et al.} \cite{jammer}\cite{jammer1} employed cooperative jamming to obtain covert
communication. To achieve the transmission of $\mathcal{O}(n)$ bits covertly to Bob over $n$ uses
of the channel, they added a ``jammer'' to the environment to help Alice for security objectives.
Soltani \emph{et al.} \cite{jammer2}\cite{DBLP:journals/corr/abs-1709-07096} considered a network
scenario where there are multiple ``friendly'' nodes that can generate interference to hide the
transmission from multiple adversaries. They assumed that the friendly nodes are in collusion with
Alice and can determine the closest node to each warden.

\begin{figure}
\centering \epsfig{file=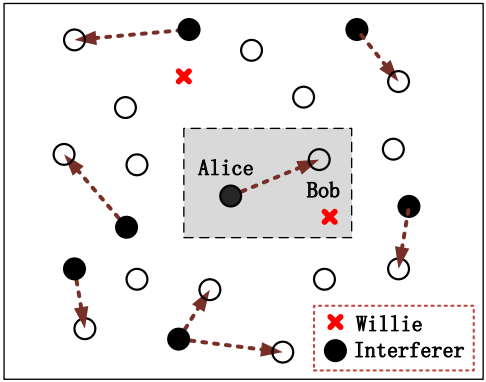, height=2in} \caption{System configuration of covert wireless communication. Alice
wishes to transmit reliably and covertly to Bob. The interferers, or other transmitters (represented by black circles) are distributed according to a two-dimensional PPP
in the presence of warden Willie (represented by red cross).}\label{deployment}
\end{figure}

In this work, we consider a large-scale wireless network, where the locations of potential
transmitters form a stationary Poisson point process (PPP), and their transmission decisions are
made independently (as depicted in Fig. \ref{deployment}). In this scenario, Bob and Willie not
only experience noise, but also interference signal from other transmitters simultaneously. Since
the measure uncertainty of aggregated interference is greater than the background noise, the
uncertainty of Willie will increase along with the increase of interference. Although the other
transmitters do not collaborate with Alice, and Bob's noise increases as well (multiuser
interference cancellation technique \cite{cancellation} is not used), we find that the covert
communication between Alice and Bob is still possible. Alice can reliably and covertly transmit
$\mathcal{O}(\log_2\sqrt{n})$ bits to Bob in $n$ channel uses when the distance between Alice and
Willie $d_{a,w}=\omega(n^{\delta/2})$ ($\delta=2/\alpha$ is the stability exponent). Although the
covert throughput is lower than the square root law and the friendly jamming scheme, its spatial
throughput is higher, and Alice does not presuppose the location knowledge of Willie. From the
perspective of network, all transmitters in the network can achieve the same covert throughput with
the same transmit power, and the larger transmit power level does not increase the probability of
being detected since Willie will also experience a stronger interference. Willie cannot determine
which node is transmitting except he can approach very close to a certain node (in this occasion
the node will find Willie and stop transmitting). ``The sound and the fury'' of the noisy wireless
channels make the network a ``shadow'' network to Willie.

\emph{Contributions}. This paper makes the following contributions:
\begin{enumerate}
  \item We considered covert wireless communications in a network scenario, and established the
      bound on reliable covert bits that may be transmitted. We found that the random
      interference in a large-scale wireless network makes the network a ``shadow network'' to
      Willie, and can achieve a high spatial throughput.
  \item Leveraging on analysis and simulation results, we proposed practical methods to improve
      the performance of covert communications in noisy wireless networks.
\end{enumerate}

The rest of the paper is structured as follows. We formulate the problem and system model in
Section \ref{ch_2}. Next, we study the covert communication with interference uncertainty in
Section \ref{ch_3}. We then present the discussions in Section \ref{ch_4} and conclude our work in
Section \ref{ch_5}.

\section{Problem Formulation and System Model}\label{ch_2}
In this section, prior to presenting the system model, we give a running example to illustrate the
problem of covert wireless communications discussed in this paper.

\subsection{Motivating Scenario}
Covert communication has a very long history. It is always related with steganography
\cite{Steganography} which conceals messages in audio, visual or textual content. However,
steganography is an application layer communication technique and is not suitable in physical-layer
covert communication. The well-known physical-layer covert communication is spread spectrum which
is using to protect wireless communication from jamming and eavesdropping \cite{Spread_Spectrum}.
Another kind of covert communications is network covert channels
\cite{covert_channel_1}\cite{covert_channel_2} in computer networks. While steganography requires
some form of content as cover, the network covert channels require network protocols as carrier. In
this paper, we consider physical-layer covert communication that employs the background noise and
the aggregated interference in wireless channels to hide transmission attempts.

Let us take the source location privacy protection in the Panda-Hunter game \cite{panda_hunter} as
an example. In the Panda-Hunter Game, a sensor network with a large number of sensors has been
deployed to monitor the habitat of pandas. As soon as a panda is observed by a sensor, this sensor
will store the observation data, and then report the observations to a sink via multi-hop wireless
channels. However, there is a hunter (the adversary Willie) in the network who tries to capture the
panda. The hunter does not care the readings of sensors, what he really cares is the location of
the message originator. To find the message originator near the panda, he listens to a sensor in
his vicinity to determine whether this sensor is transmitting message. If he finds a transmitter,
he then searches for the next sensor who is communicating with this transmitter. Via this method,
he can trace back the routing path until he reaches the message originator and catches the panda.
As a result, the source location information becomes critical and must be protected  in this
occasion.

To tackle this problem, Kamat \emph{et al.} proposed phantom routing techniques to provide
source-location privacy from the perspective of network routing \cite{panda_hunter}. Phantom
routing techniques achieve location privacy by combining flooding and single-path routing together.
From another point of view, the physical-layer covert communication can provide another kind of
solution to the Panda-Hunter game. If we can hide the transmission from the hunter in noise and
interference of the noisy wireless channels, the hunter will not able to determine which sensor is
transmitting, and therefore cannot trace back to the source. What the hunter sees is a noisy and a
shadow wireless network.

\subsection{Channel Model}
Consider a wireless communication scene where Alice (A) wishes to transmit a message to the
receiver Bob (B). Right next to them, a warden Willie (W) is eavesdropping over the wireless
channel and trying to find whether or not Alice is transmitting.

We adopt the wireless channel model similar to
\cite{square_law}\cite{DBLP:journals/corr/abs-1709-07096}, and throughout this paper we use the
similar notations. We consider a time-slotted system where the time is divided into successive
slots with equal duration. All wireless channels are assumed to suffer from discrete-time AWGN with
real-valued symbols. Alice transmits $n$ real-valued symbols $s_1^{(a)}, s_2^{(a)}, ...,
s_n^{(a)}$. The receiver Bob observes the vector $y^{(b)}_1, y^{(b)}_2, ..., y^{(b)}_n$, where
$y^{(b)}_i = s_i^{(a)}+z^{(b)}_i$, and $z^{(b)}_i$ is the noise Bob experiences which can be
expressed as $z^{(b)}_i = z^{(b)}_{i,0} + I^{(b)}_i$, where $\{z^{(b)}_{i,0}\}_{i=1}^n$ are
independent and identically distributed (i.i.d.) random variables (RVs) representing the background
noise of Bob with $z^{(b)}_{i,0}\sim \mathcal{N}(0, \sigma^2_{b,0})$, and $\{I^{(b)}_i\}^n_{i=1}$
are i.i.d. RVs characterizing the aggregated interference from other transmitters in the wireless
network.

As to Willie, he observes the vector $y^{(w)}_1, y^{(w)}_2, ..., y^{(w)}_n$, where $y^{(w)}_i =
s_i^{(a)}+z^{(w)}_i$, and $z^{(w)}_i$ is the noise Willie experiences which can be expressed as
$z^{(w)}_i = z^{(w)}_{i,0} + I^{(w)}_i$, where $\{z^{(w)}_{i,0}\}_{i=1}^n$ are i.i.d. RVs
representing the background noise of Willie with $z^{(w)}_{i,0}\sim \mathcal{N}(0,
\sigma^2_{w,0})$, and $\{I^{(w)}_i\}^n_{i=1}$ are i.i.d. RVs characterizing the aggregated
interference Willie experiences.

Suppose each node in the network is equipped with one antenna, and Bob and Willie experience the
same background noise power, i.e., $\sigma^2_{b,0} = \sigma^2_{w,0}$. Besides, different from the
occasion discussed in \cite{DBLP:journals/corr/abs-1709-07096}, no location information of Willie
and other transmitters is available in our model.

\subsection{Network Model}
Consider a large-scale wireless network, where the locations of transmitters form a stationary
Poisson point process (PPP)\cite{Haenggi_PPP} $\Pi=\{X_i\}$ on the plane $\mathbb{R}^2$. The
density of the PPP is represented by $\lambda$, denoting the average number of transmitters per
unit area. Suppose each potential transmitter has an associated receiver, the transmission
decisions are made independently across transmitters and independent of their locations for each
transmitter, and the transmission power employed for each node are constant power $P_t$. Any other
channel models with power control or threshold scheduling will have similar results with some scale
factors. Suppose the wireless channel is modeled by large-scale fading with path loss exponent
$\alpha$ ($\alpha > 2$). Let the Euclidean distance between node $i$ and node $j$ is denoted as
$d_{i,j}$. For simplicity, let the channel gain $\mathbf{h}_{i,j}$ of channel between $i$ and $j$
is static over the signaling period, and all links experience unit mean Rayleigh fading. Then, the
aggregated interference seen by Bob and Willie are the functional of the underlying PPP
$\Pi=\{X_i\}$ and the channel gain,
\begin{eqnarray}
  I_i^{(b)}  &\equiv & \sum_{k\in\Pi}\sqrt{\frac{P_t}{d_{b,k}^{\alpha}}} \mathbf{h}_{b,k}\cdot s_i^{(k)}\sim \mathcal{N}(0,\sigma^2_{I_b}) \label{eq_1}\\
  I_i^{(w)}  &\equiv & \sum_{k\in\Pi}\sqrt{\frac{P_t}{d_{w,k}^{\alpha}}} \mathbf{h}_{w,k}\cdot s_i^{(k)}\sim \mathcal{N}(0,\sigma^2_{I_w}) \label{eq_2}
\end{eqnarray}
where each $s_i^{(k)}$ is a Gaussian random variable $\mathcal{N}(0,1)$ which represents the signal
of the $k$-th transmitter in $i$-th channel use, and
\begin{eqnarray}
\sigma^2_{I_b}&=&\sum_{k\in\Pi}\frac{P_t}{d_{b,k}^{\alpha}}|\mathbf{h}_{b,k}|^2=\sum_{k\in\Pi}\frac{P_t}{d_{b,k}^{\alpha}}\Psi_{b,k}, \\
\sigma^2_{I_w}&=&\sum_{k\in\Pi}\frac{P_t}{d_{w,k}^{\alpha}}|\mathbf{h}_{w,k}|^2=\sum_{k\in\Pi}\frac{P_t}{d_{w,k}^{\alpha}}\Psi_{w,k}  \label{eq_4_4}
\end{eqnarray}
are shot noise (SN) process, representing the powers of the interference that Bob and Willie
experience, respectively. The Rayleigh fading assumption implies $\Psi_{i,j}=|\mathbf{h}_{i,j}|^2$
is exponentially distributed with $\mathbb{E}[\Psi_{i,j}] = 1$.

The powers of aggregated interferences, $\sigma^2_{I_w}$ and $\sigma^2_{I_b}$, are RVs which are
determined by the randomness of the underlying PPP of transmitters and the fading of wireless
channels. Therefore they are difficult to be predicted. Besides, the closed-form distribution of
the interference is hard to obtain and we have to bound it.

\subsection{Hypothesis Testing}
To find whether Alice is transmitting or not, Willie has to distinguish between the following two
hypotheses,
\begin{eqnarray}
\mathbf{H_0}&:& y_i^{(w)}  =  I_i^{(w)} + z_{i,0}^{(w)} \\
\mathbf{H_1}&:& y_i^{(w)}  =  \sqrt{\frac{P_t}{d_{a,w}^\alpha}}\mathbf{h}_{a,w}\cdot s_i + I_i^{(w)}
+ z_{i,0}^{(w)}
\end{eqnarray}

Based on the received vector $\mathbf{y}=(y_1^{(w)},..., y_n^{(w)})$, Willie should make a decision
on whether the received signal is noise+interference or signal plus noise+interference. We assume
that Willie employs a radiometer as his detector, and does the following statistic test
\begin{equation}
    T(\mathbf{y})=\frac{1}{n}\mathbf{y}^H\mathbf{y}=\frac{1}{n}\sum^n_{k=1}y_k^{(w)}*y_k^{(w)}>\gamma
\end{equation}
where $\gamma$ denotes Willie's detection threshold and $n$ is the number of samples.

Let $D_0$ and $D_1$ be the events that the received signal of Willie is noise+interference and
Alice's signal plus noise+interference, respectively, then the probability of false alarm and
missed detection can be denoted as $\mathbb{P}_{FA} = \mathbb{P}\{D_1 |\mathbf{H_0}\}$ and
$\mathbb{P}_{MD} = \mathbb{P}\{D_0 |\mathbf{H_1}\}$, respectively. Willie wishes to minimize his
probability of error $\mathbb{P}_e^{(w)} = (\mathbb{P}_{FA}+\mathbb{P}_{MD})/2$, but Alice's
ultimate objective is to guarantee that the average probability of error
$\mathbf{E}[\mathbb{P}_e^{(w)}] = \mathbf{E}[\mathbb{P}_{FA}+\mathbb{P}_{MD}]/2>1/2-\epsilon$ for
an arbitrarily small positive $\epsilon$.

First of all, Willie has to estimate the power level of noise+interference. The noise
$z^{(w)}_{i,0}$ not only comes from the thermal noise in his receiver but also the environmental
noise from his surroundings. Besides, the aggregated interference $I_i^{(w)}$ he sees is a random
variable which is determined by the randomness of the underlying PPP of transmitters and the
channel gains. The only way for Willie to estimate the noise+interference level is to gather
samples. However, he cannot determine definitely whether the samples he collected contain Alice's
transmission signal or not.

Besides, Alice should guarantee that the transmission is reliable, i.e., the desired receiver (Bob)
can decode her message with arbitrarily low average probability of error $\mathbb{P}^{(b)}_e$ at
long block lengths. For any $\epsilon > 0$, Bob can achieve $\mathbb{P}^{(b)}_e < \epsilon$ as
$n\rightarrow\infty$.

In this paper, we use standard Big-$\mathcal{O}$, Little-$\omega$, and Big-$\Theta$ notations to
describe bounds on asymptotic growth rates. The parameters and notation used in this paper are illustrated in Table \ref{tab_1}.

\begin{table}
\caption{parameters and notation}
\begin{tabular}{|l|l|}
  \hline
  $\mathbf{Symbol}$ & $\mathbf{Meaning}$ \\
  \hline
  $P_t$ & Transmit power  \\
  \hline
  $n$ & Number of channel use  \\
  \hline
    $\alpha$ & Path loss exponent \\
  \hline
  $\delta=2/\alpha$ & Stability exponent \\
  \hline
  $\Pi=\{X_i\}$ & PPP of potential transmitters \\
  \hline
  $\lambda$ & Intensity of PPP $\Pi$ \\
    \hline

  $s_i^{(a)}$ & Alice's signal in $i$-th channel use  \\
  \hline
  $s_i^{(k)}$ & Signal of node $k\in\Pi$ in $i$-th channel use  \\
  \hline
  $z_{i,0}^{(b)}, ~z_{i,0}^{(w)}$ & (Bob's, Willie's) background noise in $i$-th channel use \\
  \hline
  $\sigma_{b,0}^2, ~\sigma_{w,0}^2$ & Power of noise (Bob, Willie) observes  \\
  \hline
  $I_{i}^{(b)}, ~I_{i}^{(w)}$ & Interference (Bob, Willie) observes in $i$-th channel use \\
  \hline
  $\sigma_{I_b}^2, ~\sigma_{I_w}^2$ & Power of interference (Bob, Willie) observes \\
  \hline
  $\sigma_{b}^2, ~\sigma_{w}^2$ & Power of noise plus interference (Bob, Willie) observes \\
  \hline
  $d_{i,j}$ & Distance between $i$ and $j$ \\
  \hline
  $\mathbf{h}_{i,j}$ & Channel gain of channel between $i$ and $j$ \\
  \hline
  $\Psi_{i,j}$ & \tabincell{l}{ $\Psi_{i,j}=|\mathbf{h}_{i,j}|^2$ is exponentially distributed with \\ $\mathbf{E}[\Psi_{i,j}] = 1$} \\
  \hline
  $\mathcal{N}(\mu,\sigma^2)$ & \tabincell{l}{Gaussian distribution with mean $\mu$ and  variance $\sigma^2$} \\
  \hline
  $\mathbb{P}_{FA}$ &  Probability of false alarm \\
  \hline
  $\mathbb{P}_{MD}$ &  Probability of missed detection \\
  \hline
  $\mathbf{E}[X]$ & Mean of random variable $X$ \\
   \hline
   $\mathbf{Var}[X]$ & Variance of random variable $X$ \\
   \hline
   $q(\lambda)$ & Outage probability for a typical receiver \\
   \hline
   $\tau(\lambda)$ & Spatial throughput of successful transmissions \\
   \hline
\end{tabular}\label{tab_1}
\end{table}

\section{Covert Communication With Interference Uncertainty in Noisy Wireless Networks}\label{ch_3}
In this section, we first present a theorem on the amount of information that can be transmitted
covertly and reliably over AWGN channels in a noisy wireless network, then present its
achievability and converse proof.

\textbf{Theorem 1}. \emph{Suppose a large-scale wireless network, where transmission decisions of
nodes are made randomly, and the locations of transmitters form a PPP on the plane $\mathbb{R}^2$.
When the distance between Alice and Willie $d_{a,w}=\omega(n^{\delta/4})$, Alice can covertly and
reliably transmit $\mathcal{O}(\log_2\sqrt{n})$ bits to Bob in $n$ channel uses in the case that
$\alpha=4$ ($\delta=2/\alpha$ is the stability exponent). Conversely, if the distance
$d_{a,w}=\mathcal{O}(n^{\delta/4})$, and Alice attempts to send $\omega(\log_2\sqrt{n})$ bits to
Bob in $n$ channel uses, then, as $n\rightarrow\infty$, either Willie can detect her transmission
with arbitrarily low probability of error $\mathbb{P}^{(w)}_e$, or Bob cannot decode Alice's
message with arbitrarily low error probability $\mathbb{P}^{(b)}_e$.}


\subsection{Achievability}
To transmit messages to Bob reliably, Alice should encode her messages. In this paper, we use the
classical encoder scheme used in \cite{square_law} and suppose that Alice and Bob have a shared
secret of sufficient length. At first, Alice and Bob leverage the shared secret and random coding
arguments to generate a secret codebook. Then Alice's channel encoder takes as input message of
length $L$ bits and encodes them into codewords of length $n$ at the rate of $R = L/n$ bits/symbol.
Each codeword is a zero-mean Gaussian random $\mathcal{N}(0,P_t)$ where $P_t$ is the transmit
power.

\subsubsection{Covertness}
Alice's objective is to hide her transmission attempts from being detected by Willie. If Willie's
probability of error $\mathbf{E}[\mathbb{P}_e^{(w)}] =
\mathbf{E}[\mathbb{P}_{FA}+\mathbb{P}_{MD}]/2>1/2-\epsilon$ for an arbitrarily small positive
$\epsilon$, then we can say that the covertness is satisfied.

Different from the cases studied in \cite{square_law}\cite{DBLP:journals/corr/abs-1709-07096},
Alice and Bob are located in a noisy wireless network. No location information of Willie and other
potential transmitters is available, and Alice cannot collude with other ``friendly'' nodes. Willie
not only experiences the background noise, but also the aggregated interference from other
transmitters in the network. Therefore the power of noise and interference Willie experiences can
be expressed as
\begin{equation}\label{eq_3}
   \sigma_w^2=\sigma^2_{w,0}+\sigma_{I_w}^2,
\end{equation}
where $\sigma^2_{w,0}$ is the power of the background noise, $\sigma_{I_w}^2$ is the power of the
aggregated interference from other transmitters (defined in Equ. (\ref{eq_4_4})). In general, the
interference is more difficult to be predicted than the background noise, since the randomness of
aggregated interference comes from the randomness of PPP $\Pi$ and the fading channels, especially
in a mobile wireless network.

Let $\mathbb{P}_0$ be the joint probability density function (PDF) of $\mathbf{y}=(y_1^{(w)},...,
y_n^{(w)})$ when $\mathbf{H_0}$ is true, $\mathbb{P}_1$ be the joint PDF of $\mathbf{y}$ when
$\mathbf{H_1}$ is true. Using the same analysis methods and the results from
\cite{square_law}\cite{DBLP:journals/corr/abs-1709-07096},
if Willie employs the optimal hypothesis test to minimize his probability of detection error
$\mathbb{P}_e^{(w)}$, then
\begin{equation}\label{eq_3_1}
   \mathbb{P}_e^{(w)}\geq\frac{1}{2}-\sqrt{\frac{1}{8}D(\mathbb{P}_1||\mathbb{P}_0)},
\end{equation}
where $D(\mathbb{P}_1||\mathbb{P}_0)$ is the relative entropy between $\mathbb{P}_1$ and
$\mathbb{P}_0$, and the lower bound $\mathbb{P}_e^{(w)}$ can be expressed as follows
\begin{eqnarray}\label{eq_4}
  \mathbb{P}_e^{(w)} &\geq & \frac{1}{2}-\sqrt{\frac{n}{8}}\cdot\frac{P_t\Psi_{a,w}}{2\sigma_w^2d^{\alpha}_{a,w}} \nonumber\\
      &=&  \frac{1}{2}-\sqrt{\frac{n}{8}}\cdot\frac{P_t\Psi_{a,w}}{2d^{\alpha}_{a,w}}\cdot\frac{1}{\sigma^2_{w,0}+\sigma_{I_w}^2} \nonumber\\
      &\geq & \frac{1}{2}-\sqrt{\frac{n}{8}}\cdot\frac{P_t\Psi_{a,w}}{2d^{\alpha}_{a,w}}\cdot\frac{1}{\sigma_{I_w}^2}.
\end{eqnarray}
The last step is due to $\sigma^2_{w,0}<<\sigma_{I_w}^2$, since in a dense and large-scale wireless
network, the background noise is negligible compared to the aggregated interference from other
transmitters \cite{limited}. Then the mean of $\mathbb{P}_e^{(w)}$ is
\begin{eqnarray}\label{eq_8}
    \mathbf{E}[\mathbb{P}_e^{(w)}]&\geq& \frac{1}{2}-\sqrt{\frac{n}{8}}\cdot\frac{P_t\mathbf{E}[\Psi_{a,w}]}{2d^{\alpha}_{a,w}}\cdot\mathbf{E}\biggl[\frac{1}{\sigma_{I_w}^2}\biggr] \nonumber\\
    &=& \frac{1}{2}-\sqrt{\frac{n}{8}}\cdot\frac{P_t}{2d^{\alpha}_{a,w}}\cdot\mathbf{E}\biggl[\frac{1}{\sigma_{I_w}^2}\biggr]
\end{eqnarray}
for all links experience unit mean Rayleigh fading.

To estimate $\mathbf{E}[1/\sigma_{I_w}^2]$, we should have the closed-form expression of the
distribution of $\sigma^2_{I_w}=\sum_{k\in\Pi}\frac{P_t}{d_{w,k}^{\alpha}}\Psi_{w,k}$. However,
$\sigma^2_{I_w}$ is an RV whose randomness originates from the random positions in PPP $\Pi$ and
the fading channels. It obeys a stable distribution without closed-form expression for its PDF or
cumulative distribution function (CDF). To address wireless network capacity, Weber \emph{et al.}
\cite{Weber} employed tools from stochastic geometry to obtain asymptotically tight bounds on the
distribution of the signal-to-interference (SIR) level in a wireless network, yielding tight bounds
on its complementary cumulative distribution function (CCDF). Next we leverage the bounds on CCDF
to estimate the expectation $\mathbf{E}[1/\sigma_{I_w}^2]$.

Define a random variable
\begin{equation}
\mathbf{Y} = \frac{\Sigma_{i\in\Pi}P_t\Psi_{i,w}d^{-\alpha}_{i,w}}{P_t\Psi_{a,w}d^{-\alpha}_{a,w}}=\frac{\sigma^2_{I_w}}{P_t\Psi_{a,w}d^{-\alpha}_{a,w}},
\end{equation}
then, the lower bound on the CCDF of RV $\mathbf{Y}$, $\bar{F}^l_\mathbf{Y}(y)$, can be expressed
as \cite{Weber},
\begin{equation}
\bar{F}_\mathbf{Y}^l(y)=\kappa\lambda y^{-\delta} + \mathcal{O}(y^{-2\delta}),
\end{equation}
where $\kappa = \pi\mathbf{E}[\Psi^\delta]\mathbf{E}[\Psi^{-\delta}]\mathbf{E}[d^2_{a,w}]$, $\lambda$ is
the intensity of attempted transmissions in PPP $\Pi$, and $\delta=2/\alpha$. When $\Psi \sim
\mathrm{Exp}(1)$, $\kappa =
\pi\Gamma(1+\delta)\Gamma(1-\delta)d^2_{a,w}=\frac{\pi^2\delta}{\sin(\pi\delta)}d^2_{a,w}$.

Therefore the upper bound of CDF of $\mathbf{Y}$ can be represented as
\begin{equation}
F_\mathbf{Y}^u(y)=1-\kappa\lambda y^{-\delta}.
\end{equation}

Next we can get the upper bound of CDF of $\sigma^2_{I_w}$ as
\begin{eqnarray}\label{eq_5}
  F^u_{\sigma^2_{I_w}}(x) &=& \mathbb{P}\{\sigma^2_{I_w}<x\} = \mathbb{P}\{P_t\Psi_{a,w}d_{a,w}^{-\alpha}\mathbf{Y}<x\} \nonumber\\
   &=& \mathbb{P}\{\mathbf{Y}<\frac{x}{P_t\Psi_{a,w}d_{a,w}^{-\alpha}}\} \nonumber\\
   &=& 1-\kappa\lambda\beta^\delta x^{-\delta}
\end{eqnarray}
where $\beta=P_t\Psi_{a,w}d_{a,w}^{-\alpha}$. For simplicity, we assume the channel gain of channel
between Alice and Willie is static and constant, $\mathbf{h}_{a,w}=1$. Then $\beta$ can be denoted
as $\beta=P_t d_{a,w}^{-\alpha}$.

Therefore the upper bound of PDF of $\sigma^2_{I_w}$ can be represented as
\begin{equation}\label{eq_CDF}
f^u_{\sigma^2_{I_w}}(x)=\kappa\lambda\beta^\delta\delta x^{-(\delta+1)}, ~~x\in [(\kappa\lambda)^{1/\delta}\beta, +\infty).
\end{equation}
where we set $x\in [(\kappa\lambda)^{1/\delta}\beta, +\infty)$ to normalize the function so that it
describes a probability density.

Given the upper bound of PDF of $\sigma^2_{I_w}$, we can upper bound $\mathbf{E}[1/\sigma_{I_w}^2]$
as follows
\begin{eqnarray}\label{eq_14}
  \mathbf{E}\biggl[\frac{1}{\sigma_{I_w}^2}\biggr] &\leq & \int^{\infty}_{(\kappa\lambda)^{1/\delta}\beta}\kappa\lambda\beta^\delta\delta x^{-(\delta+1)}\cdot\frac{1}{x}\mathrm{d}x \nonumber\\
   &=& \frac{\delta}{\delta+1}(\kappa\lambda)^{-1/\delta}\beta^{-1}
\end{eqnarray}

Thus, (\ref{eq_8}) and (\ref{eq_14}) yield the lower bound of $\mathbf{E}[\mathbb{P}_e^{(w)}]$ as
\begin{eqnarray}\label{eq_15}
    \mathbf{E}[\mathbb{P}_e^{(w)}] &\geq & \frac{1}{2}-\sqrt{\frac{n}{8}}\cdot\frac{P_t}{2d^{\alpha}_{a,w}}\cdot\mathbf{E}\biggl[\frac{1}{\sigma_{I_w}^2}\biggr] \nonumber\\
         &\geq & \frac{1}{2}-\sqrt{\frac{n}{8}}\cdot\frac{P_t}{2d^{\alpha}_{a,w}}\cdot\frac{\delta}{\delta+1}(\kappa\lambda)^{-1/\delta}\beta^{-1} \nonumber\\
         & = & \frac{1}{2}-\sqrt{\frac{n}{8}}\cdot\frac{P_t}{2d^{\alpha}_{a,w}}\cdot\frac{\delta}{\delta+1}\lambda^{-1/\delta}\biggl(\frac{\pi^2\delta}{\sin\pi\delta}\biggr)^{-1/\delta} \nonumber\\
         & & \times d_{a,w}^{-2/\delta}P_t^{-1}d^{\alpha}_{a,w} \nonumber\\
         & = & \frac{1}{2}-\sqrt{\frac{n}{8}}\cdot\frac{\delta}{2(\delta+1)}\biggl(\frac{\pi^2\delta\lambda}{\sin\pi\delta}\biggr)^{-1/\delta}\cdot d_{a,w}^{-2/\delta}
\end{eqnarray}

Suppose $\mathbf{E}[\mathbb{P}_e^{(w)}]\geq \frac{1}{2}-\epsilon$ for any $\epsilon>0$, then we
should set
\begin{equation}
\sqrt{\frac{n}{8}}\cdot\frac{\delta}{2(\delta+1)}\biggl(\frac{\pi^2\delta\lambda}{\sin\pi\delta}\biggr)^{-1/\delta}\cdot d_{a,w}^{-2/\delta}<\epsilon.
\end{equation}
Let $c = \sqrt{1/8}\cdot\frac{\delta}{2(\delta+1)}(\frac{\pi^2\delta}{\sin\pi\delta})^{-1/\delta}$,
we have
\begin{equation}
d_{a,w}>(c/\epsilon)^{\delta/2}n^{\delta/4}.
\end{equation}

Therefore, as long as $d_{a,w}=\omega(n^{\delta/4})$, we can get
$\mathbf{E}[\mathbb{P}_e^{(w)}]\geq \frac{1}{2}-\epsilon$ for any $\epsilon>0$. This implies that
there is no limitation on the transmit power $P_t$ of Alice and other potential transmitters, the
critical factor is the distance between Alice and Willie.  This result is different from the works
of Bash \cite{square_law} and Soltani \cite{DBLP:journals/corr/abs-1709-07096}, in which Alice's
symbol power is a decreasing function of the codeword length $n$. While this may appear
counter-intuitive, the result in fact is explicable. We believe the reasons are two folds. First,
higher transmission signal power will create larger interference which will make Willie more
difficult to judge. Secondly, more close to the transmitter will give Willie more accurate
estimation. This theoretical result is also verified using the experimental results in Section
\ref{ch_4}.

\subsubsection{Reliability}
Next, we estimate Bob's decoding error probability, denoted by $\mathbb{P}^{(b)}_e$. Let the noise
power that Bob experiences be
\begin{equation}
\sigma^2_b=\sigma^2_{b,0}+\sigma^2_{I_b}
\end{equation}
where $\sigma^2_{b,0}$ is the power of background noise Bob observes, $\sigma_{I_b}^2$ is the power
of the aggregated interference from other transmitters in the network. By utilizing the same
approach in \cite{square_law}, Bob's decoding error probability can be lower bounded as follows,
\begin{eqnarray}
  \mathbb{P}^{(b)}_e(\sigma^2_b) &\leq& 2^{nR-\tfrac{n}{2}\log_2\bigl(1+\tfrac{P_t}{2\sigma^2_b}\bigr)} \nonumber\\
   &=& 2^{nR-\tfrac{n}{2}\log_2\bigl[1+\tfrac{P_t}{2(\sigma^2_{b,0}+\sigma^2_{I_b})}\bigr]} \nonumber\\
   &=& 2^{nR}\biggl[1+\frac{P_t}{2(\sigma^2_{b,0}+\sigma^2_{I_b})}\biggr]^{-n/2} \nonumber\\
   &\leq & 2^{nR}\biggl[1+\frac{P_t}{2(\sigma^2_{b,0}+\sigma^2_{I_b})}\frac{n}{2}\biggr]^{-1}
\end{eqnarray}
where the last step is obtained by the following inequality
\cite{DBLP:journals/corr/abs-1709-07096}
\begin{equation}
(1+x)^{-r}\leq (1+rx)^{-1} ,~~\text{for any}~~ r\geq 1~~ \text{and}~~ x>-1.
\end{equation}

Hence the upper bound of Bob's average decoding error probability can be estimated as follows
\begin{eqnarray}\label{eq_21}
  \mathbf{E}[\mathbb{P}^{(b)}_e(\sigma^2_b)] &\leq& \mathbf{E}\biggl[2^{nR}\biggl(1+\frac{nP_t/4}{\sigma^2_{b,0}+\sigma^2_{I_b}}\biggr)^{-1}\biggr] \nonumber\\
   &< & \int_0^{\infty}2^{nR}\biggl(1+\frac{nP_t/4}{\sigma^2_{b,0}+x}\biggr)^{-1}f^u_{\sigma^2_{I_b}}(x)\mathrm{d}x \nonumber\\
   &=& 2^{nR}\int_{(\kappa\lambda)^{\tfrac{1}{\delta}}\beta}^{\infty}\biggl(1+\frac{nP_t/4}{\sigma^2_{b,0}+x}\biggr)^{-1} \nonumber\\
    & & \times \kappa\lambda\beta^\delta\delta x^{-(\delta+1)} \mathrm{d}x
\end{eqnarray}
where $f^u_{\sigma^2_{I_b}}(x)$ is the upper bound of PDF of $\sigma^2_{I_b}$ which obeys the
similar distribution as $\sigma^2_{I_w}$ (Equ. (\ref{eq_CDF})),
\begin{equation}\label{eq_CDF1}
f^u_{\sigma^2_{I_b}}(x)=\kappa\lambda\beta^\delta\delta x^{-(\delta+1)}, ~~x\in [(\kappa\lambda)^{1/\delta}\beta, +\infty).
\end{equation}
where $\beta=P_t\Psi_{a,b}d_{a,b}^{-\alpha}$.

Although the interference Bob and Willie observe obey the similar distribution, they are
correlative random variables. This is because the interference is caused by common randomness of
the PPP $\Pi$ \cite{Interference_Haenggi}. When Bob is far away from Willie, the correlation
between $\sigma^2_{I_b}$ and $\sigma^2_{I_w}$ is almost zero, which implies that the interferences
seen by Bob and Willie are approximately independent. When Bob and Willie are very close to each
other, they experience almost the same interference. In this occasion, $\sigma^2_{I_b}$ and
$\sigma^2_{I_w}$ are approximately identical random variables.

Let $a = nP_t/4$, the path loss exponent $\alpha=4$, then $\delta=1/2$. The Equ. (\ref{eq_21}) can
be calculated as follows
\begin{eqnarray}\label{eq_22}
  \mathbf{E}[\mathbb{P}^{(b)}_e(\sigma^2_b)] &<& 2^{nR}\int_{(\kappa\lambda)^{\tfrac{1}{\delta}}\beta}^{\infty}\biggl(1+\frac{a}{\sigma^2_{b,0}+x}\biggr)^{-1} \nonumber\\
    & & \times \kappa\lambda\beta^\delta\delta x^{-(\delta+1)} \mathrm{d}x  \nonumber\\
    &=& 2^{nR}\kappa\lambda\beta^\delta\delta\biggl[\frac{\pi a}{(a+\sigma^2_{b,0})^{3/2}}- \\
    & & \frac{2a\arctan\tfrac{\kappa\lambda\beta^\delta}{\sqrt{a+\sigma^2_{b,0}}}}{(a+\sigma^2_{b,0})^{3/2}}+\frac{2\sigma^2_{b,0}}{\kappa\lambda\beta^\delta(a+\sigma^2_{b,0})}\biggr] \nonumber
\end{eqnarray}

As $\beta=P_t\Psi_{a,b}d_{a,b}^{-\alpha}$,
$\kappa=\frac{\pi^2\delta}{\sin(\pi\delta)}d_{a,b}^2=\frac{\pi^2}{2}d_{a,b}^2$ for $\delta=1/2$,
when $n$ is large enough, we have
\begin{equation}
  a = nP_t/4\gg \sigma^2_{b,0}, ~~ a+\sigma^2_{b,0}\approx a.
\end{equation}
and
\begin{eqnarray}
\frac{2\sigma^2_{b,0}}{\kappa\lambda\beta^\delta(a+\sigma^2_{b,0})} &\rightarrow& 0 \\
\frac{2a\arctan\tfrac{\kappa\lambda\beta^\delta}{\sqrt{a+\sigma^2_{b,0}}}}{(a+\sigma^2_{b,0})^{3/2}}   &>& \frac{2\sigma^2_{b,0}}{\kappa\lambda\beta^\delta(a+\sigma^2_{b,0})}   \\
\frac{\pi a}{(a+\sigma^2_{b,0})^{3/2}} &\rightarrow& \frac{\pi}{\sqrt{a}}
\end{eqnarray}
Therefore we have
\begin{eqnarray}
  \mathbf{E}[\mathbb{P}^{(b)}_e(\sigma^2_b)] &<& 2^{nR}\kappa\lambda\beta^\delta\delta\frac{\pi a}{(a+\sigma^2_{b,0})^{3/2}} \nonumber\\
   &<& 2^{nR}\kappa\lambda\beta^\delta\delta\frac{2\pi}{\sqrt{nP_t}} \nonumber\\
   &=& 2^{nR}\frac{\pi^2}{2}d_{a,b}^2\lambda P_t^{1/2}\mathbf{E}[\Psi^{1/2}]d_{a,b}^{-\alpha/2}\delta\frac{2\pi}{\sqrt{nP_t}} \nonumber\\
   &=& 2^{nR}\frac{\pi^{7/2}\lambda\delta}{2\sqrt{n}}
\end{eqnarray}
where $\mathbf{E}[\Psi^{1/2}]=\Gamma(1+1/2)=\sqrt{\pi}/2$ for $\Psi\sim \mathrm{Exp}(1)$.

Let $\mathbf{E}[\mathbb{P}^{(b)}_e(\sigma^2_b)]\leq\epsilon$ for any $\epsilon>0$, we have
\begin{equation}
  nR\leq \log_2\biggl(\frac{2\epsilon}{\pi^{7/2}\lambda\delta}\cdot\sqrt{n}\biggr),
\end{equation}
which implies that Bob can receive
\begin{equation}
L=\mathcal{O}(\log_2\sqrt{n})~~ \text{bits}
\end{equation}
reliably in $n$ channel uses in the case that $\alpha=4$. This may be a pessimistic result at first
glance since it is much lower than the bound derived in the work of Bash \cite{square_law}, i.e.,
Bob can reliably receive $\mathcal{O}(\sqrt{n})$ bits in $n$ channel uses. This is reasonable
because Bob experiences not only the background noise but also the aggregated interference,
resulting lower transmit throughput. However, in the work of Bash, Alice's symbol power is a
decreasing function of the codeword length $n$, i.e., her average symbol power
$P_f\leq\frac{cf(n)}{\sqrt{n}}$. When Bob use threshold-scheduling scheme to receive signal, Bob
will have higher outage probability as $n\rightarrow\infty$. This is because Alice's symbol power
will become very lower to ensure the covertness as $n\rightarrow\infty$. If we hide communications
in noisy wireless networks, the spatial throughput is higher than the work of Bash in which only
background noise is considered. This will be discussed in Section \ref{ch_4}.

\subsection{Converse}
In this subsection we present the converse of the Theorem. Suppose Willie make a decision on
whether the received signal includes Alice's signal based on the received vector
$\mathbf{y}=(y_1^{(w)},..., y_n^{(w)})$. He computes
$T(\mathbf{y})=\frac{1}{n}\mathbf{y}^H\mathbf{y}=\frac{1}{n}\sum^n_{k=1}y_k^{(w)}*y_k^{(w)}$, and
employs a radiometer as his detector to do the following statistical test with $\gamma$ as his
detection threshold,
\begin{eqnarray}
    \text{If}~T(\mathbf{y})&<&\sigma^2_w+\gamma, ~~\text{Willie accepts $\mathbf{\mathbf{H_0}}$} \nonumber\\
    \text{If}~T(\mathbf{y})&\geq&\sigma^2_w+\gamma, ~~\text{Willie accepts $\mathbf{\mathbf{H_1}}$}
\end{eqnarray}
where $\sigma^2_w$ is the power of noise Willie experiences (defined in Equ. (\ref{eq_3})), and we
assume that Willie knows $\sigma^2_w$.

When $\mathbf{H_0}$ is true, $y_i^{(w)}=z^{(w)}_{i,0}+I^{(w)}_i$, where
$z^{(w)}_{i,0}\sim\mathcal{N}(0,\sigma^2_{w,0})$ is the background noise, and $I^{(w)}_i$
represents the aggregated interference from other transmitters (defined in Equ. (\ref{eq_2})). The
transmitter $k\in\Pi$ sends codewords $s_i^{(k)}$ in the $i$-th channel use. Willie observes
\begin{equation}\label{eq_H0}
    y_i^{(w)}\sim \mathcal{N}(I^{(w)}_i,\sigma^2_{w,0}) = \mathcal{N}\biggl(\sum_{k\in\Pi}\sqrt{\frac{P_t}{d_{w,k}^{\alpha}}}\mathbf{h}_{w,k} s_i^{(k)},\sigma^2_{w,0}\biggr)
\end{equation}
which contains readings of mean-shifted noise.

Next we estimate the mean and variance of $T(\mathbf{y})$. At first, we have to compute the mean
and variance of $y_i^{(w)}$. Because the RV
$Z=\bigl(\frac{y_i^{(w)}-I^{(w)}_i}{\sigma_{w,0}}\bigr)^2\sim\chi^2(1)$, its mean and variance are
1 and 2, respectively. Hence,
\begin{eqnarray}
   & & \mathbf{E}\biggl[\biggl(\frac{y_i^{(w)}-I^{(w)}_i}{\sigma_{w,0}}\biggr)^2\biggr] \nonumber\\
   &=& \frac{1}{\sigma^2_{w,0}}\biggl(\mathbf{E}[(y_i^{(w)})^2]-2\mathbf{E}[y_i^{(w)}I^{(w)}_i]+(I^{(w)}_i)^2\biggr) \nonumber\\
   &=&  \frac{1}{\sigma^2_{w,0}}\biggl(\mathbf{E}[(y_i^{(w)})^2]-(I^{(w)}_i)^2\biggr)  = 1
\end{eqnarray}
yields $\mathbf{E}[(y_i^{(w)})^2]=\sigma^2_{w,0}+(I^{(w)}_i)^2$. Given this, the mean of
$T(\mathbf{y})$ can be computed as
\begin{eqnarray}
  \mathbf{E}[T(\mathbf{y})|\mathbf{H_0}] &=& \mathbf{E}\biggl[\frac{1}{n}\sum^n_{k=1}y_k^{(w)}*y_k^{(w)}\biggr]= \mathbf{E}[(y_i^{(w)})^2]  \nonumber\\
   &=& \sigma^2_{w,0}+\mathbf{E}[(I^{(w)}_i)^2] \nonumber\\
   &=& \sigma^2_{w,0} + \sigma^2_{I_w}.
\end{eqnarray}
The last equation comes from the fact that
$\mathbf{E}[(I^{(w)}_i)^2]=\mathbf{Var}[I^{(w)}_i]+(\mathbf{E}[I^{(w)}_i])^2=\sigma^2_{I_w}$ where
$\sigma^2_{I_w}=\sum_{k\in\Pi}\frac{P_t}{d_{w,k}^{\alpha}}\Psi_{w,k}$.

Because RVs $(y_i^{(w)})^2$ and $y_i^{(w)}$ are uncorrelated random variables, the variance of
$T(\mathbf{y})$ can be computed in the same method as follows
\begin{eqnarray}
   & &  \mathbf{Var}\biggl[\biggl(\frac{y_i^{(w)}-I^{(w)}_i}{\sigma_{w,0}}\biggr)^2\biggr] \nonumber\\
  &=& \frac{1}{\sigma^4_{w,0}}\biggl(\mathbf{Var}[(y_i^{(w)})^2]-4(I^{(w)}_i)^2\mathbf{Var}[y_i^{(w)}]\biggr) \nonumber\\
   &=&  \frac{1}{\sigma^4_{w,0}}\biggl(\mathbf{Var}[(y_i^{(w)})^2]-4(I^{(w)}_i)^2\sigma^2_{w,0}\biggr)  = 2
\end{eqnarray}
and $\mathbf{Var}[(y_i^{(w)})^2]=2\sigma^4_{w,0}+4(I^{(w)}_i)^2\sigma^2_{w,0}$. Hence the variance
of $T(\mathbf{y})$ can be estimated as follows
\begin{eqnarray}
  \mathbf{Var}[T(\mathbf{y})|\mathbf{H_0}] &=& \mathbf{Var}\biggl[\frac{1}{n}\sum^n_{k=1}y_k^{(w)}*y_k^{(w)}\biggr]= \frac{\mathbf{Var}[(y_i^{(w)})^2]}{n}  \nonumber\\
   &=& \frac{1}{n}\biggl(2\sigma^4_{w,0}+4\mathbf{E}[(I^{(w)}_i)^2]\sigma^2_{w,0}\biggr) \nonumber\\
   &=& \frac{1}{n}\biggl(2\sigma^4_{w,0}+4\sigma^2_{I_w}\sigma^2_{w,0}\biggr). \label{eq_H0_D}
\end{eqnarray}

When $\mathbf{H_1}$ is true, Alice transmits a codeword signal which is included in the signal $\mathbf{y}$
that Willie observes. In this occasion, Willie observes
\begin{eqnarray}\label{eq_H1}
    y_i^{(w)} &\sim& \mathcal{N}\biggl(\sqrt{\frac{P_t}{d_{w,a}^{\alpha}}}\mathbf{h}_{w,a}s_i+I^{(w)}_i,~\sigma^2_{w,0}\biggr) \\
        &\sim& \mathcal{N}\biggl(\sqrt{\frac{P_t}{d_{w,a}^{\alpha}}}\mathbf{h}_{w,a}s_i+\sum_{k\in\Pi}\sqrt{\frac{P_t}{d_{w,k}^{\alpha}}} \mathbf{h}_{w,k} s_i^{(k)},~\sigma^2_{w,0}\biggr) \nonumber
\end{eqnarray}

Then using the similar method we can derive the following results,
\begin{eqnarray}
  \mathbf{E}[T(\mathbf{y})|\mathbf{H_1}] &=& \sigma^2_{w,0} + \frac{P_t}{d_{a,w}^\alpha} + \sigma^2_{I_w}  \label{eq_H1_E}  \\
  \mathbf{Var}[T(\mathbf{y})|\mathbf{H_1}] &=& \frac{1}{n}\biggl[2\sigma^4_{w,0}+4\biggl(\frac{P_t}{d_{a,w}^\alpha}+\sigma^2_{I_w}\biggr)\sigma^2_{w,0}\biggr]. \label{eq_H1_D}
\end{eqnarray}

By Chebyshev's inequality, the probability $\mathbb{P}_{FA}$ can be bounded as follows
\begin{eqnarray}
  \mathbb{P}_{FA} &=& \mathbb{P}\{T(\mathbf{y})>\sigma^2_w+\gamma\} \nonumber\\
   &=&  \mathbb{P}\{T(\mathbf{y})>\sigma^2_{w,0}+\sigma^2_{I_w}+\gamma\} \nonumber\\
   &\leq & \mathbb{P}\{|T(\mathbf{y})-(\sigma^2_{w,0}+\sigma^2_{I_w})|>\gamma\} \nonumber\\
   &\leq & \frac{\mathbf{Var}[T(\mathbf{y})|\mathbf{H_0}]}{\gamma^2} \nonumber\\
    &=& \frac{1}{n\gamma^2}(2\sigma^4_{w,0}+4\sigma^2_{I_w}\sigma^2_{w,0})
\end{eqnarray}
and
\begin{equation}\label{eq_37}
    \mathbf{E}[\mathbb{P}_{FA}]\leq \frac{1}{n\gamma^2}\biggl(2\sigma^4_{w,0}+4\mathbf{E}[\sigma^2_{I_w}]\sigma^2_{w,0}\biggr).
\end{equation}

Next we need to estimate the mean of $\sigma^2_{I_w}$ which is the aggregated interference and is a
functional of the underlying PPP $\Pi$. However, its mean is not exist if we employ the unbounded
path loss law (this may be partly due to the singularity of the path loss law at the origin). We
then use a modified path loss law to estimate the mean of $\sigma^2_{I_w}$,
\begin{equation}\label{eq_law}
    l(r)\equiv r^{-\alpha}\mathbf{1}_{r\geq\rho},~~r\in\mathbb{R}_+, ~~\text{for}~\rho\geq 0.
\end{equation}

This law truncates around the origin and thus removes the singularity of impulse response function
$l(r)\equiv r^{-\alpha}$. The guard zone around the receiver (a ball of radius $\rho$) can be
interpreted as assuming any two nodes can't get too close. Strictly speaking, transmitters no
longer form a PPP under this bounded path loss law, but a hard-core point process in this case. For
relatively small guard zones, this model yields rather accurate results. For $\rho>0$, the mean and
variance of $\sigma^2_{I_w}$ are finite and can be given as \cite{Interference_Haenggi}
\begin{eqnarray}
  \mathbf{E}[\sigma^2_{I_w}] &=& \frac{\lambda d c_d}{\alpha-d}\mathbf{E}[\Psi]\mathbf{E}[P_t]\rho^{d-\alpha} \\
  \mathbf{Var}[\sigma^2_{I_w}] &=& \frac{\lambda d c_d}{2\alpha-d}\mathbf{E}[\Psi^2]\mathbf{E}[P_t^2]\rho^{d-2\alpha}
\end{eqnarray}
where $d$ is the spatial dimension of the network, the relevant values of $c_d$ are: $c_1=2$,
$c_2=\pi$, $c_3=4\pi/3$.

When $d=2$, $\alpha=4$, constant transmit power $P_t$, and the fading $\Psi\sim\mathrm{Exp}(1)$, we
have
\begin{equation}
\mathbf{E}[\sigma^2_{I_w}]=\frac{\pi\lambda}{\rho^2}\cdot P_t
\end{equation}
and
\begin{equation}\label{eq_40}
    \mathbf{E}[\mathbb{P}_{FA}]\leq \frac{1}{n\gamma^2}\biggl(2\sigma^4_{w,0}+\frac{4\pi\lambda}{\rho^2}P_t\sigma^2_{w,0}\biggr).
\end{equation}

For any $\epsilon>0$, Willie can set his threshold
\begin{equation}
\gamma=\frac{\sigma^2_{w,0}}{\sqrt{n\epsilon}}\sqrt{\frac{4\pi\lambda}{\rho^2}P_t+2\sigma^2_{w,0}}
\end{equation}
to satisfy $\mathbf{E}[\mathbb{P}_{FA}]\leq\epsilon$.

Because the background noise is negligible compared to the aggregated interference from other
transmitters in a dense wireless network, $P_t\gg\sigma^2_{w,0}$, given
$c=2\sqrt{\pi\lambda}\sigma^2_{w,0}/\rho$, Willie can set its detection threshold to
\begin{equation}\label{eq_44}
\gamma=\Theta\biggl(c\sqrt{\frac{P_t}{n}}\biggr).
\end{equation}

Next the $\mathbb{P}_{MD}$ can be upper bounded for the given detection threshold $\gamma$ in
Equ.(\ref{eq_44}) as follows
\begin{eqnarray}
  \mathbb{P}_{MD} &=& \mathbb{P}\{T(\mathbf{y})<\sigma^2_w+\gamma\} \nonumber\\
   &\leq & \mathbb{P}\biggl\{\biggl|T(\mathbf{y})-\biggl(\sigma^2_w+\frac{P_t}{d^\alpha_{a,w}}\biggr)\biggr|>\frac{P_t}{d^\alpha_{a,w}}-\gamma\biggr\} \nonumber\\
   &\leq & \frac{1}{(\frac{P_t}{d^\alpha_{a,w}}-\gamma)^2}\mathbf{Var}[T(\mathbf{y})|\mathbf{H_1}]
\end{eqnarray}
and its mean can be estimated as
\begin{equation}
  \mathbf{E}[\mathbb{P}_{MD}] \leq \frac{1}{(\frac{P_t}{d^\alpha_{a,w}}-\gamma)^2}\frac{1}{n}\biggl[2\sigma^4_{w,0}+4\biggl(\frac{P_t}{d_{a,w}^\alpha}+\mathbf{E}[\sigma^2_{I_w}]\biggr)\sigma^2_{w,0}\biggr].
\end{equation}

Next we assume $\alpha=4$, $\delta=2/\alpha=1/2$. Since
$\gamma=\Theta\bigl(c\sqrt{\frac{P_t}{n}}\bigr)$,
$\mathbf{E}[\sigma^2_{I_w}]=\frac{\pi\lambda}{\rho^2}P_t$, then if
$d_{a,w}=\Theta(n^{\delta/4})=\Theta(n^{1/8})$, Willie can upper bound
$\mathbf{E}[\mathbb{P}_{MD}]$ as follows
\begin{eqnarray}
  \mathbf{E}[\mathbb{P}_{MD}]
  &\leq& \frac{2\sigma^2_{w,0}}{\bigl(\frac{P_t}{\sqrt{n}}-c\sqrt{\frac{P_t}{n}}\bigr)^2}\frac{1}{n}\biggl[\sigma^2_{w,0}+2\biggl(\frac{P_t}{\sqrt{n}}+\frac{\pi\lambda}{\rho^2}P_t\biggr)\biggr] \nonumber\\
  &=& \frac{2\sigma^2_{w,0}}{(\sqrt{P_t}-c)^2}\biggl(\frac{\sigma^2_{w,0}}{P_t}+\frac{2}{\sqrt{n}}+\frac{2\pi\lambda}{\rho^2}\biggr).
\end{eqnarray}


Consequently, when $n\rightarrow\infty$ and $P_t\gg\sigma^2_{w,0}$, we have
\begin{equation}
\frac{2}{\sqrt{n}}\rightarrow 0,~~ \frac{\sigma^2_{w,0}}{P_t}\rightarrow 0,
\end{equation}
and
\begin{equation}
\mathbf{E}[\mathbb{P}_{MD}] \leq \frac{4\pi\lambda\sigma^2_{w,0}}{\rho^2}\frac{1}{(\sqrt{P_t}-c)^2},
\end{equation}
which implies that in the case $d_{a,w}=\Theta(n^{\delta/4})$, when
$P_t\geq\bigl(\sqrt{\frac{4\pi\lambda\sigma^2_{w,0}}{\epsilon'\rho^2}}+c\bigr)^2$, then
\begin{equation}
\mathbf{E}[\mathbb{P}_{MD}]\leq\epsilon' ~~\text{for any} ~~ \epsilon'>0.
\end{equation}

Hence Alice cannot covertly send any codeword with arbitrary transmit power $P_t$ when the distance
is $d_{a,w}=\mathcal{O}(n^{\delta/4})$. To avoid being detected by Willie, Alice must be certain
that there is no eavesdropper in her immediate vicinity. In the case that
$d_{a,w}=\mathcal{O}(n^{\delta/4})$, she cannot transmit with arbitrary transmit power to achieve a
higher covert transmission rate than $\mathcal{O}(\log_2\sqrt{n})$.

\section{Discussions}\label{ch_4}
\subsection{Spatial Throughput}
The spatial throughput is the expected spatial density of successful transmissions in a wireless
network \cite{Weber}
\begin{equation}
\tau(\lambda)=\lambda(1-q(\lambda))
\end{equation}
where $q(\lambda)$ denotes the probability of transmission outage when the intensity of attempted
transmissions is $\lambda$ for given SINR requirement $\xi$.

In the work of Bash \emph{et al.} \cite{square_law}, only background noise is taken into account,
Alice can transmit $\mathcal{O}(\sqrt{n})$ bits reliably and covertly to Bob over $n$ uses of the
AWGN wireless channel. To achieve the covertness, Alice must set her average symbol power
$P\leq\frac{cf(n)}{\sqrt{n}}$. Soltani \emph{et
al.}\cite{jammer2}\cite{DBLP:journals/corr/abs-1709-07096} further expanded the work of Bash. They
introduced the friendly node closest to Willie to produce artificial noise. They showed that this
method allows Alice to reliably and covertly send
$\mathcal{O}(\min\{n,\lambda^{\alpha/2}\sqrt{n}\})$ bits to Bob in $n$ channel uses when there is
only one adversary. In their network settings, $\lambda$ is the density of friendly nodes on the
plane $\mathbb{R}^2$, and Alice must set her average symbol power
$P_a=\mathcal{O}(\frac{c\lambda^{\alpha/2}}{\sqrt{n}})$ to avoid being detected by Willie. Thus,
given an SINR threshold $\xi$, $\sigma_{b,0}^2\geq 1$, and Rayleigh fading with
$\Psi\sim\mathrm{Exp}(1)$, the outage probability  of Soltani's method is
\begin{eqnarray}
  q^J(\lambda) &=& \mathbb{P}\biggl\{\mathrm{SINR}=\frac{P_a\Psi d^{-\alpha}_{a,b}}{\sigma^2_{b,0}+P_f\Psi d^{-\alpha}_{a,f}}<\xi\biggr\} \nonumber \\
    &\geq&  \mathbb{P}\{P_a\Psi d^{-\alpha}_{a,b}<\xi\} \nonumber \\
    &\geq& \mathbb{P}\biggl\{\frac{c\lambda^{\alpha/2}}{\sqrt{n}}\Psi d^{-\alpha}_{a,b}<\xi\biggr\} \nonumber \\
    &=& \mathbb{P}\biggl\{\Psi<\frac{1}{c\lambda^{\alpha/2}}d^{\alpha}_{a,b}\xi\sqrt{n}\biggr\} \nonumber \\
    &=& 1-\exp\biggl\{-\frac{1}{c\lambda^{\alpha/2}}d^{\alpha}_{a,b}\xi\sqrt{n}\biggr\}
\end{eqnarray}
where $P_f$ is the jamming power of the friendly node, and $d_{a,f}$ is the distance between Alice
and the friendly node. Then the spatial throughput of the network is
\begin{equation}\label{eq_spatial_J}
\tau^J(\lambda)=\lambda(1-q^J(\lambda))\leq\lambda\exp\biggl\{-\frac{1}{c\lambda^{\alpha/2}}d^{\alpha}_{a,b}\xi\sqrt{n}\biggr\}.
\end{equation}

If we hide communications in the aggregated interference of a noisy wireless network with
randomized transmissions in Rayleigh fading channel and the SINR threshold is set to $\xi$, the
spatial throughput is \cite{Weber}
\begin{equation}\label{eq_spatial_I}
\tau^I(\lambda)=\lambda\exp\{-\pi\lambda\xi^{\delta}d^2_{a,b}\Gamma(1+\delta)\Gamma(1-\delta)\}
\end{equation}
where $\delta=2/\alpha$.

As a result of Equ. (\ref{eq_spatial_J}) and (\ref{eq_spatial_I}), we can state that, by using a
friendly jammer near Willie to help Alice, Alice can reliably and covertly send
$\mathcal{O}(\min\{n,\lambda^{\alpha/2}\sqrt{n}\})$ bits to Bob in $n$ channel uses, which is
higher than $\mathcal{O}(\log_2\sqrt{n})$ bits when the aggregated interference is involved. But as
$n\rightarrow\infty$, the spatial throughput of the jamming scheme $\tau^J(\lambda)$ reduces to
zero, and the covert communication hiding in interference can achieve a constant spatial throughput
$\tau^I(\lambda)$ which is higher than $\tau^J(\lambda)$. Hence, this approach, while has lower
covert throughput for any pair of nodes, has a considerable higher throughput from the network
perspective.

\subsection{Interference Uncertainty}
From the analysis above, we found that the interference can indeed increase the privacy throughput.
If we can deliberately deploy interferers to further increase the interference Willie experiences
and not harm Bob, the security performance can be enhanced, such as the methods discussed in
\cite{jammer}\cite{jammer1}\cite{jammer2}.

\begin{figure}
\centering \epsfig{file=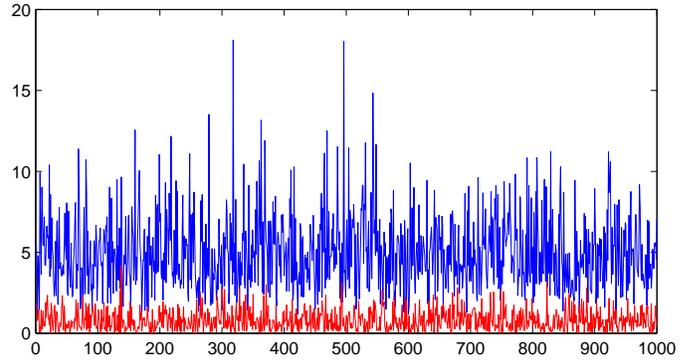, height=1.9in}
\caption{Sequences of 1000 realizations of noise and aggregated interference. Here a bounded path
loss law is used, $l(x)=\frac{1}{1+\parallel x\parallel^\alpha}$. The transmit power $P_t$ of nodes are all unity,
links experience unit mean Rayleigh fading, $\Psi\sim\mathrm{Exp}(1)$, and $\alpha=4$. A reference point is located at the center of a square area 100m$\times$100m.
Interferers deployed in this area form a PPP on the plane $\mathbb{R}^2$ with $\lambda=1$.
Interference the reference point sees is depicted in blue, the noise is depicted in red.}\label{Interference}
\end{figure}

Overall, the improvement comes from the increased interference uncertainty. If there is only noise
from Willie's surroundings, he may estimate the noise level by gathering samples although the
background noise can be unpredictable to some extent. However, the aggregated interference is more
difficult to be predicted, since the randomness of interference comes from the randomness of PPP
$\Pi$ and the fading channels. Fig.\ref{Interference} illustrates this situation by sequences of
realizations of the noise (Normal distribution with the variance one) and the aggregated
interference. From the figure, we find that the interference has greater dispersion than the
background noise, thus it is more difficult to sample interferences to obtain a proper interference
level.

Additionally, the aggregated interference is always dominated by the interference generated by the
nearest interferer. If an interferer gets closer to Willie than Alice, Willie will be overwhelmed
by the signal of the interferer, and his decision will be uncertain. Let $r_1$ be the distance of
the nearest interferer of Willie, $f_{R_1}(r)$ be the PDF of the nearest-neighbor distance
distribution on the plane $\mathbb{R}^2$ \cite{distances}, then
\begin{eqnarray}
  \mathbb{P}\{r_1<d_{a,w}\} &=& \int_0^{d_{a,w}}f_{R_1}(r)\mathrm{d}r \nonumber \\
   &=& \int_0^{d_{a,w}}2\pi\lambda r\exp(-\pi\lambda r^2)\mathrm{d}r \nonumber \\
   &=& 1-\exp(-\pi\lambda d_{a,w}^2).
\end{eqnarray}
We see that when $d_{a,w}=1$ and $\lambda=1$, $\mathbb{P}\{r_1<d_{a,w}\} = 0.9568$ - that is, there
is a dramatically high probability that Willie will experience more interference from the nearest
interferer. He will confront a dilemma to make a binary decision. In a dense and noisy wireless
network, Willie cannot determine which node is actually transmitting if he cannot get closer than
$\Theta(n^{2/\delta})$ and cannot sure no other nodes located in his detect region.

\subsection{Practical Method and Experimental Results}
In the proof of Theorem 1, when Willie samples the noise to determine the threshold of his detector
(radiometer), we presuppose that Willie knows whether Alice is transmitting or not, and he knows
the power level of $\sigma^2_{I_w}$. In practice, Willie has no prior knowledge on whether Alice
transmits or not during his sampling process. This implies that Willie's sample $y^{(w)}_i$ follows
the distribution
\begin{equation}\label{eq_H111}
    y_i^{(w)} \sim \mathcal{N}\biggl(\sqrt{\frac{P_t}{d_{w,a}^{\alpha}}}\mathbf{h}\cdot s_i\mathbf{1}_A+\sum_{k\in\Pi}\sqrt{\frac{P_t}{d_{w,k}^{\alpha}}} \mathbf{h}\cdot s_i^{(k)},\sigma^2_{w,0}\biggr),
\end{equation}
where $\mathbf{1}_A$ is an indicator function, $\mathbf{1}_A=1$ when Alice is transmitting,
$\mathbf{1}_A=0$ when Alice is silent, and the transmission probability
$\mathbb{P}\{\mathbf{1}_A=1\}=p$.

If Alice can transmit messages and be silent alternately, Willie cannot be certain whether the $n$
samples contain Alice's signals or not. To confuse Willie, Alice should not generate burst traffic,
but transforming the bulk message into a smooth network traffic with transmission and silence
alternatively. She can divide the time into slots, then put message into small packets. After that,
Alice sends a packet in a time slot and keeps silence for the next slot, and so on. Via this
scheduling scheme, Alice can guarantee that Willie's samples are the mix of noise and signal which
are undistinguishable by Willie.

\begin{figure}
\centering \epsfig{file=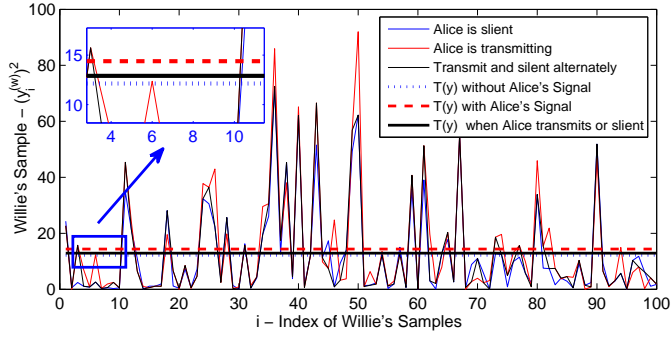, height=1.8in}
\caption{Sequences of 100 Willie's samples $[y_1^{(w)}]^2,..., [y_n^{(w)}]^2$ in the cases that Alice is silent, transmitting, or transmit and silent alternately. $T(\mathbf{y})=\frac{1}{n}\sum^n_{k=1}[y_k^{(w)}]^2$ in three cases are depicted as three lines. Here a bounded path loss law is used, $l(x)=\min\{1, r^{-\alpha}\}$. The transmit power $P_t$ is unity,
links experience unit mean Rayleigh fading, $\Psi\sim\mathrm{Exp}(1)$, $\alpha=4$, and $\sigma^2_{w,0}=1$. Willie is located at the center of a square area 100m$\times$100m. The distance between Alice and Willie $d_{a,w}=1$. Interferers deployed in this area form a PPP on the plane $\mathbb{R}^2$ with $\lambda=1$.
}\label{Willie_samples}
\end{figure}

Next we provide an experimentally-supported analysis of this methods. Fig. \ref{Willie_samples}
illustrates an example of sequences of 100 Willie's samples $[y_1^{(w)}]^2,..., [y_n^{(w)}]^2$ in
the case that Alice is silent, transmitting, or transmitting and silent alternately. Willie then
computes $T(\mathbf{y})=\frac{1}{n}\sum^n_{k=1}[y_k^{(w)}]^2$. Clearly, when Alice alternates
transmission with silence, Willie's sample value $T(\textbf{y})$ will decrease and quite near the
value when Alice is silent. For this reason, the alternation of Alice can increase Willie's
uncertainty. The transmitted signals resemble white noise, and are sufficiently weak in this way.

\begin{figure}
\centering \epsfig{file=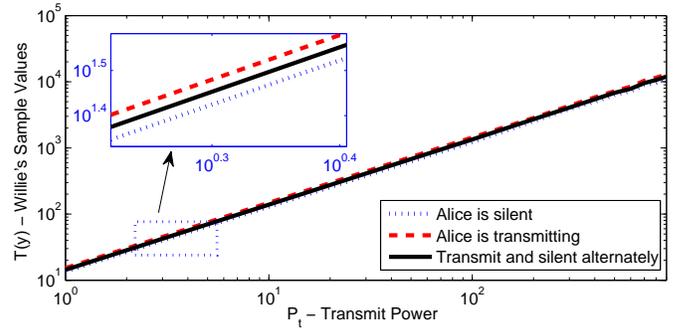, height=1.8in}
\caption{The transmit power $P_t$ versus Willie's sample values $T(\textbf{y})$ which are the average of 100 experiment runs, each with the number of samples $n=500$.
During each run of the simulation, to obtain a sample $y_i^{(w)}$, a random wireless network obeying PPP on the plane $\mathbb{R}^2$ is generated.
Here the distance between Alice and Willie $d_{a,w}=1$.}\label{pt_y_mean}
\end{figure}

With the same simulation settings of Fig. \ref{Willie_samples}, we evaluate Willie's sample values
$T(\textbf{y})$ by varying the transmit power $P_t$. As displayed in Fig. \ref{pt_y_mean}, the
value $T(\textbf{y})$ changing with $P_t$ is displayed in three cases, i.e., Alice is transmitting,
silent, as well as transmitting and silent alternately. We find that when Alice employs the
alternation method, Willie's sample values decrease, approximating to the case Alice is silent.
Further, the results indicate that higher transmit power cannot lead to stronger capability for
Willie to distinguish Alice's transmission behavior. With the transmit power increases, Willie's
sample values $T(\textbf{y})$ increase. However, the aggregated interference increases as well,
resulting in the gap of sample values between Alice's transmission and silence does not increase.
Consequently, this is consistent with the result of Theorem 1, which indicates that increasing the
transmit power $P_t$ does not increase the risk of being detected by Willie.

\begin{figure}
\centering \epsfig{file=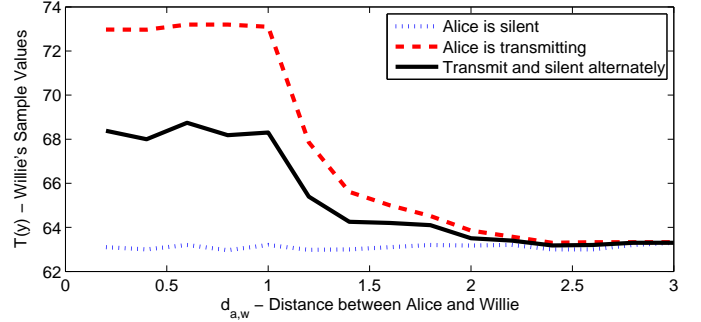, height=1.8in}
\caption{The distance between Alice and Willie $d_{a,w}$ versus Willie's sample values $T(\textbf{y})$ which are the average of 100 experiment runs,
each with the number of samples $n=500$. Here the transmit power $P_t=10$, and the transmission
probability $p=0.5$.}\label{daw_y_mean_1}
\end{figure}

Further, as Theorm 1 states, one of critical factors affecting covert communication is the
parameter $d_{a,w}$, the distance between Alice and Willie, which should satisfies
$d_{a,w}=\omega(n^{\delta/4})$ to ensure communication covertly. Fig. \ref{daw_y_mean_1}
illustrates Willie's sample values $T(\textbf{y})$ by varying the distance $d_{a,w}$. As the
results show, when Alice is silent, Willie's sample values $T(\textbf{y})$ barely change since
Willie only experiences the noise and aggregated interference. When Alice is transmitting,
persistence or alteration, Willie's sample values increase with decreasing the distance $d_{a,w}$.
When $d_{a,w}\leq 1$, Willie's sample values become relatively stable since we employ the bounded
path loss law $l(x)=\min\{1, r^{-\alpha}\}$.

\begin{figure} \centering
\subfigure[n=1000]{ \epsfig{file=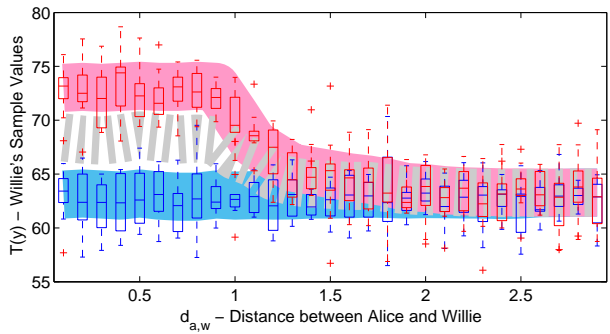, height=1.8in }}
\subfigure[n=3000]{ \epsfig{file=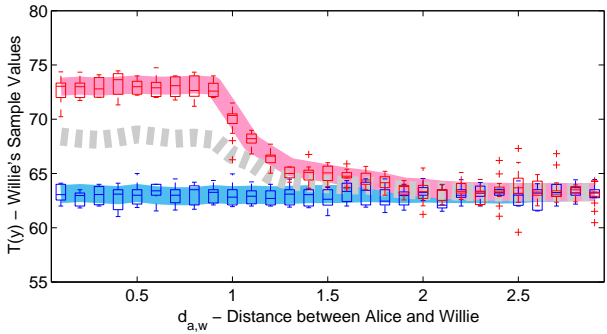, height=1.8in} }
\caption{The discreteness of Willie's sample values $T(\textbf{y})$ versus the distance $d_{a,w}$ when the number of samples $n=1000$ and $n=3000$.
At each subfigure, three simulation curves are given (from top to bottom): Alice is transmitting, transmitting and silent alternately (with transmission
probability $p=0.5$), and silent completely.
For each occasion, given a value $d_{a,w}$ and $n$, we implement 20 experiment runs to obtain 20 sample values $T(\textbf{y})$, and depict the discreteness of $T(\textbf{y})$ in boxplot form.
The width of curves also represent the dispersion degree of Willie's sample values.}
\label{boxplot}
\end{figure}

For the following analysis, we evaluate the effect of the number of samples $n$ on the distance
between Alice and Willie $d_{a,w}$. We start by comparing Willie's sample values by varying $n$ to
show the difference in performance. The results in Fig. \ref{boxplot} shows $T(\textbf{y})$ with
respect to the distance $d_{a,w}$ when $n=1000$ and $n=3000$. As can be seen, although the curves
of the average $T(\textbf{y})$ do not change, the discreteness of $T(\textbf{y})$ decreases with
increasing the number of samples $n$. As to Willie, to detect Alice's transmission attempts, he
should distinguish the three lines in the picture with relatively low probability of error. The
only way to decrease the probability of error is increasing the number of samples. By choosing a
larger value for $n$, Willie's uncertainty on noise and interference decreases, hence he can stay
far away from Alice to detect her transmission attempt. As illustrated in Fig. \ref{boxplot}(a),
Willie cannot distinguish Alice's transmission from silence when he stays at a distance of more
than 1 meter to Alice. However, when Willie increases the number of samples, he can distinguish
Alice's behavior far away. As depicted in Fig. \ref{boxplot}(b), Willie can detect Alice's
transmission at the distance between 1 and 1.5 meters with low probability of error. Overall, this
experimental result agrees with the theoretical derivation and conclusion of Theorem 1, i.e., given
the value $n$, the distance between Alice and Willie should be larger than a bound to ensure the
covertness, and the bound of $d_{a,w}$ increases with the increasing of $n$,
$d_{a,w}=\omega(n^{\delta/4})$.

\subsection{Traffic Shaping and Willie's Sample Dilemma}
\begin{figure}
\centering \epsfig{file=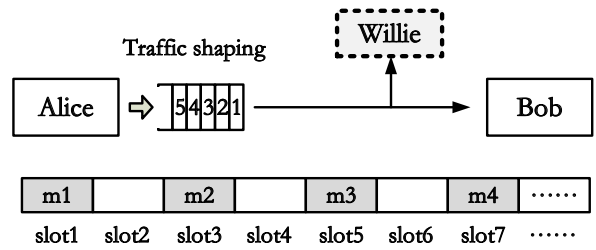, height=1.3in}
\caption{Traffic shaping.}\label{Traffic}
\end{figure}

A practical way for Alice to avoid being detected is leveraging traffic shaping \cite{rfc2475} as
her transmission scheduling such that packet transmission and silence are alternative in time
slots.  As depicted in Fig. \ref{Traffic}, Alice divides a chunk of data into packets and transmit
each packet in the odd slots. Traffic shaping may be implemented with for example the leaky bucket
or token bucket algorithm. Traffic shaping used in this occasion is not to optimize or increase
usable bandwidth, it is used to uniform the transmission of Alice. Although it may increases the
transmission latency, Willie's uncertainty also increases.

\textbf{Willie's Sample Dilemma}: As discussed earlier, when Willie increases the number of
samples, he can distinguish Alice's behavior at greater distance. According to Equ. (\ref{eq_H0_D})
and (\ref{eq_H1_D}), a larger quantity of samples of Willie will lead to smaller variance of
$T(\mathbf{y})|\mathbf{H_0}$ and $T(\mathbf{y})|\mathbf{H_1}$. Hence Willie's uncertainty will
decrease with the increase of samples. However, to confuse Willie, Alice can divide the slots
sufficiently small and insert her message in the slots uniformly. When Willie gathers samples, more
samples he collects, more signals of Alice will be included in his samples. If Willie employs a
radiometer as his detector, it is difficult for him to distinguish between hypotheses
$\mathbf{H_0}$ and $\mathbf{H_1}$. Therefore, how to determine the number of samples is a dilemma
that Willie has to be confronted with. As to Alice, her better policy is alternating transmission
and idle periods, that is, ``\emph{telling a short story in a long period, speaking for a short
time and taking a rest for a while.}''

\subsection{Time Interval and Slow Start}
Because Willie employs a radiometer as his detector, to obtain a relatively accurate estimation of
noise, he first gathers a large quantity of samples to determine his detection threshold. After
that, he leverages the threshold to determine whether Alice is transmitting or not in an interval
by comparing the threshold with the samples in this new interval.

\begin{figure} \centering
\subfigure[Alice's transmission scheduling and Willie's sampling interval]{ \epsfig{file=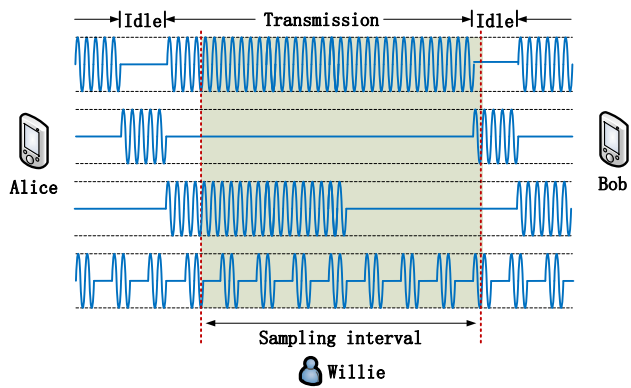, height=2in }}
\subfigure[Sampling Value]{ \epsfig{file=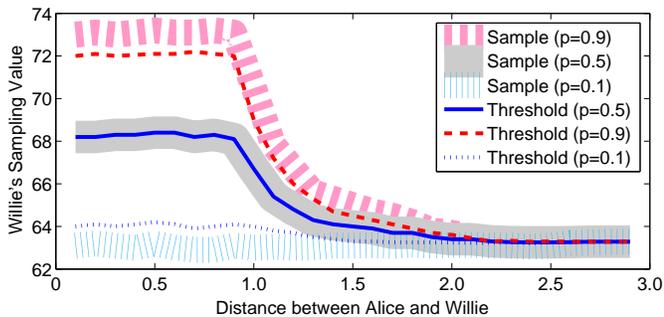, height=1.7in} }
\caption{(a) Willie's sampling interval versus Alice's transmission scheduling with the transmission probability $p=0.9$, $p=0.1$, and $p=0.5$ (from top to bottom).
(b) Willie's sampling value by varying the distance between Alice and Willie for different transmission scheduling schemes.
The threshold is obtained via a large quantity of samples. The sample is computed through the sample values in a relative short sampling interval (shown in subfigure (a)).
The width of sample curves represent the dispersion degree of Willie's sampling value. }
\label{sampling_interval}
\end{figure}

At first, Alice has to determine the proper time intervals of transmission and idle states. Fig.
\ref{sampling_interval}(a) illustrates Alice's transmission scheduling with different slot
assignments, Fig. \ref{sampling_interval}(b) depicts Willie's sampling value by varying the
distance between Alice and Willie for different transmission scheduling schemes. We can find that,
if Alice's transmission slot is wider than the idle slot, Willie's sampling value may greater than
his detection threshold with high probability, resulting in the exposure of Alice's transmission
behavior. When Alice's transmission slot is shorter than her idle slot, Willie's sampling value may
less than his detection threshold, and Willie can deduce that Alice is in her idle state at this
time. Therefore, the optimal slot assignment method is to set the length of the transmission slot
and idle slot in the same length, i.e., $p=0.5$. As shown in Fig. \ref{sampling_interval}(b) when
$p=0.5$, the sampling values change around the threshold value, making it difficult for Willie to
determine whether Alice is transmitting or not. Besides, the slot should be as short as possible to
make Willie's samples contain more Alice's signal.

\textbf{Slow Start and Slow Stop}: If Alice is transmitting while Willie is sampling the channel to
determine the interference level and his detection threshold, Willie's samples will contain Alice's
signals. If Willie later employs this threshold to test Alice's behavior, he will not be able to
ascertain Alice's transmission behavior. However, if Alice is not transmitting while Willie begins
his sampling, Willie's detection threshold will be lower than that when Alice is transmitting.
Therefore Willie will have higher probability to find the transmission attempt when Alice transmits
later. Lower transmission probability implies weaker transmitted signal. As depicted in Fig.
\ref{sampling_interval}(b), Willie's sampling value with $p=0.1$ is much lower than the sampling
value with $p=0.5$ but a little higher than that when Alice is idle. Therefore, to decrease the
probability of being detected, Alice should transmit with lower transmission probability $p=0.1$
from the very beginning, and slowly increasing until $p=0.5$. Conversely, Alice should slowly stop
her transmission in case of being detection when she has no more messages to transmit further.

In the scene that network traffic is sparse or not evenly spread in the whole network, the
aggregated interference may be too weak to cover the transmission attempts or unevenly distributed.
In the case of sparse traffic, potential transmitters should resort to recruiting ``friendly''
nodes to generate artificial noise, such as the methods used in
\cite{DBLP:journals/corr/abs-1709-07096}. In the case of uneven traffic distribution, the better
way is using some effective methods, such as routing protocols, to homogenize the network traffic.

In most practical scenarios, to detect the transmission attempt of Alice, Willie should approach
Alice as close as possible, and ensure that there is no other node located closer to Willie than
Alice. Otherwise, Willie cannot determine which one is the actual transmitter. But in a wireless
network, some wireless nodes are probably placed on towers, trees, or buildings, Willie cannot get
close enough as he wishes. Furthermore, wireless networks are diverse and complicated. If Willie is
not definitely sure that there is no other transmitter in his vicinity, he cannot ascertain that
Alice is transmitting. However, in a mobile wireless network, some mobile nodes may move into the
detection region of Willie, and increase the uncertainty of Willie. Therefore mobile can improve
the performance of covert communication to some extend.

\section{Conclusions}\label{ch_5}
In this paper, we have studied the covert wireless communication with the consideration of
interference uncertainty. Prior studies on covert communication only considered the background
noise uncertainty, or introduced collaborative jammers producing artificial noise to help Alice in
hiding the communication. By introducing interference measurement uncertainty, we find that
uncertainty in noise and interference experienced by Willie is beneficial to Alice, and she can
achieve undetectable communication with better performance. If Alice want to hide communications
with interference in noisy wireless networks, she can reliably and covertly transmit
$\mathcal{O}(\log_2\sqrt{n})$ bits to Bob in $n$ channel uses. Although the covert rate is lower
than the square root law and the friendly jamming scheme, its spatial throughput is higher as
$n\rightarrow\infty$. From the network perspective, the communications are hidden in the noisy
wireless networks. It is difficult for Willie to ascertain whether a certain user is transmitting
or not, and what he sees is merely a shadow wireless network.



\end{document}